\magnification=1200
\headline{\ifnum\pageno=1 \nopagenumbers
\else \hss\number \pageno \fi}

\overfullrule=0pt
\footline={\hfil}
\font\boldgreek=cmmib10
\textfont9=\boldgreek
\mathchardef\mypsi="0920
\def\bfpsi{{\fam=9 \mypsi}\fam=1}
\mathchardef\myphi="091E

\def\lsim{\raise0.3ex\hbox{$<$\kern-0.75em\raise-1.1ex\hbox{$\sim$}}}
\def\gsim{\raise0.3ex\hbox{$>$\kern-0.75em\raise-1.1ex\hbox{$\sim$}}}
\baselineskip=10pt
\vbox to 1,5truecm{}
\centerline{\bf J/$\bfpsi$ AND $\bfpsi$' SUPPRESSION IN HEAVY ION
COLLISIONS}\bigskip

\bigskip \centerline{by}\medskip
\centerline{{\bf A. Capella, A. Kaidalov}\footnote{*}{Permanent address : ITEP,
B.
Cheremushkinskaya 25, 117259 Moscow, Russia}{\bf, A. Kouider Akil}}
\smallskip
 \centerline{Laboratoire de Physique Th\'eorique et Hautes Energies
\footnote{**}{Laboratoire associ\'e au Centre National de la Recherche
Scientifique - URA D0063}}  \centerline{Universit\'e de Paris XI, b	timent 211,
91405
Orsay cedex, France}
\bigskip
\centerline{\bf C. Gerschel}
\smallskip
\centerline{Institut de Physique Nucl\'eaire}
\centerline{Universit\'e de Paris XI, b	timent 100,
91406 Orsay cedex, France}\bigskip \bigskip \bigskip\baselineskip=20pt
\noindent
${\bf Abstract}$ \par
We study the combined effect of nuclear absorption and final state interaction
with
co-moving hadrons on the $J/\psi$ and $\psi '$ suppression in proton-nucleus
and
nucleus-nucleus collisions. We show that a reasonable description of the
experimental
data can be achieved with theoretically meaningful values of the cross-sections
involved and without introducing any discontinuity in the $J/\psi$ or $\psi '$
survival
probabilities. \par

\vbox to 4 truecm{}

\noindent LPTHE Orsay 96-55 \par
\noindent July 1996

\vfill\supereject
In 1986 Matsui and Satz [1] proposed $J/\psi$ suppression in heavy ion
collisions as a
signal of quark gluon plasma (QGP) formation. This suppression results from
Debye
screening in a medium of deconfined quarks and gluons. Shortly afterwards,
the NA38 collaboration tested this idea and found that in $O$-$U$ and $S$-$U$
collisions the ratio $J/\psi$ over di-muon continuum decreases with increasing
centrality [2]. However, very soon, two alternative explanations involving
conventional
physics - i.e. no phase transition - were proposed. In the first one, known as
nuclear
absorption [3] [4], the $c\bar{c}$ pair wave packet produced inside the
nucleus, is
modified by nuclear collisions in such a way that it does not project into
$J/\psi$ but
into open charm. It was shown that the absorptive cross-section $\sigma_{abs}$
needed
to explain the $A$-dependence of $J/\psi$ production in $pA$ collisions also
does
explain the suppression found in nucleus-nucleus. This cross-section turns out
to be
about 6 mb. In the second explanation, known as interaction with co-movers, the
$J/\psi$ produced outside the nucleus is surrounded by a dense system of
hadrons
(mainly pions) and converts into open charm due to interactions in the medium
[5]. This
interaction takes place at low energies - which, however, have to be large
enough to
overcome threshold effects. Reliable theoretical calculations of the
corresponding
cross-section $\sigma_{co}^{\psi}$ show that it increases very slowly with
energy from
threshold [6]. In view of that, the second interpretation has been
progressively
abandoned in favor of nuclear absorption [7]. \par

Very recent data for $Pb$ $Pb$ collisions obtained by the NA50 collaboration
show an
anomalous $J/\psi$ suppression [8]. The ratio $J/\psi$ over Drell-Yan (DY) is
two times
smaller than the extrapolation of the $O$-$U$ and $SU$ data based on nuclear
absorption -
the statistical significance of this discrepancy being of nine standard
deviations. These
data provide a most exciting hint of QGP formation and some interpretations in
this
context have been  presented at the QM'96 conference [9]. In them a
discontinuity in the $J/\psi$ survival probability is assumed when some
threshold of
local energy density is reached. \par

In this note we present an attempt to describe the observed $J/\psi$
suppression using
the combined effect of nuclear absorption and interaction with co-movers -
without
introducing any discontinuity in the survival probability [10]. For the nuclear
absorption we adopt the formalism of refs. [3, 4]. For simplicity, we use the
exponential
form of the $J/\psi$ suppression given in ref. [4] and used in the experimental
papers,
namely $$S_1 = \exp (- \rho \ L \ \sigma_{abs} ) \eqno(1)$$
\noindent where $\rho = 0.138$ nucleon/fm$^3$ is the nuclear density and $L$ is
the
length of nuclear matter crossed by the $c\bar{c}$ pair. The more involved
formalism of
ref. [3] gives results very similar to those obtained from (1). \par

We turn next to the final state interaction with produced hadrons (co-movers).
A
rigourous treatment of this interaction is very complicated and is not
available in the
literature. We use the simple treatment proposed in refs. [11]. The decrease in
the
spatial density of $J/\psi$ at point $x$ due to the interaction $\psi$-$h$ is
given by
[12]
$$\Delta {dN^{\psi} \over d^4x} = \rho^{\psi} (x) \ \rho^h(x)
\sigma_{co}^{\psi}
\eqno(2)$$
\noindent with $d^4x = \tau d\tau dy d^2s$, where $\tau$ is the proper time,
$y$ the
space-time rapidity (to be later on identified with the usual rapidity) and
$d^2s$ an
element of transverse area. $\sigma_{co}^{\psi}$ is the part of the $\psi$-$h$
cross-section which does not contain the $J/\psi$ in the final state, averaged
over the
momentum distribution of the colliding particles. The effect of thresholds will
be taken
into account in an effective way through the value of $\sigma_{co}^{\psi}$.
Assuming
longitudinal boost invariance and a dilution of the densities $\rho^{\psi}$ and
$\rho^h$
of the type $1/\tau$ (i.e. neglecting transverse expansion), we get from (2)
$$\left . {dN^{\psi} \over dy} \right |_{\tau_0 + \Delta \tau} = \int d^2s
{dN^{\psi}
\over dy d^2s} {dN^h \over dy d^2s} \sigma_{co}^{\psi} \ell n \left | {\tau_0 +
\tau
\over \tau_0} \right | \eqno(3)$$
\noindent where $\tau_0$ is the formation time and $\tau$ is the duration of
the hadronic phase. \par

We want to express the densities $dN/dyd^2s$ in terms of the observables
$dN/dy$. We
have [11]
$$\int d^2s {dN^{\psi} \over dy d^2s}(b) {dN^h \over dy d^2s}(b) = G(b)
{dN^{\psi}
\over dy}(b) {dN^h \over dy}(b) \eqno(4)$$
\noindent where $b$ is the impact parameter of the collision and the
geometrical factor
$G(b)$ is given by
$$G(b) = {\int d^2s \ T_A^2(s) \ T_B^2(b -s) \over T_{AB}^2(b)} \ \ \ ,
\eqno(5)$$
\noindent which has an obvious geometrical interpretation. For the nuclear
profile
$T_A(b)$ we use standard Saxon-Woods. For the proton we use a Gaussian
profile with $R_p = 0.6$ fm. Using
(3)-(5) we have~: $$\left . {dN^{\psi} \over dy} \right |_{\tau_0 + \Delta
\tau}(b) =
\left . {dN^{\psi} \over dy} \right |_{\tau_0} (b) \left [ 1 -
\sigma_{co}^{\psi} G(b)
\ell n \left | {\tau_0 + \Delta \tau \over \tau_0} \right | {dN^h \over dy}(b)
\right ]
\eqno(6)$$ \noindent and, for a finite time interval,
$${dN^{\psi} \over dy}(b) = \left . {dN^{\psi} \over dy} \right |_{\tau_0}(b)
\exp
\left [ - \sigma_{co}^{\psi} G(b) \ell n \left | {\tau_0 + \tau (b) \over
\tau_0}
\right | {dN^h \over dy} (b) \right ] \equiv \left . {dN^{\psi} \over dy}
\right
|_{\tau_0}(b) \ S_2(b) \ . \eqno(7)$$ \noindent We have now to specify the
duration time
of the interaction $\tau (b)$. This time is not well known. We use the
following ansatz
[11]. In the case of Gaussian profiles one has
$$G = {1 \over 2 \pi} \left [ {2 \over 3} {R_A^2 \cdot R_B^2 \over R_A^2 +
R_B^2}
\right ] \eqno(8)$$
\noindent where $R$ are the rms radii. In ref. [13] it was found that the
quantity in
brackets in (8) is precisely the geometrical HBT squared transverse radius.
Following the
arguments of [14] we take it as a measure of the duration of interaction and
therefore
use
$$\tau (b) = [2 \pi G(b) ]^{-1/2} \eqno(9)$$
\noindent  For the formation time $\tau_0$ we take $\tau_0$ = 1 fm [14]. Of
course our
results depend on the value of $\tau_0$. However, this dependence can, to a
large
extent, be compensated by a small change of $\sigma_{co}^{\psi}$. Finally
$dN^h(b)/dy$ is
a measurable quantity. In order to avoid model estimates we use the
experimental value of
$E_{T}$ as a measure of the hadronic activity. More precisely, for $SU$
collisions, where
the NA38 calorimeter covers a range $- 1.3 < \eta_{cm} < 1.1$ not far from the
one of
the dimuon ($0 < \eta < 1$), we take
$${dN^h \over dy} (b) = {3 E_T(b) \over \Delta \eta
<p_T>} \ \ \ , \eqno(10)$$ \noindent where $E_T$ is the average energy of
neutrals
measured by the calorimeter in each centrality bin, $\Delta \eta = 2.4$, and
$<p_T>$ =
0.35 GeV. Note, however, that our results depend only on the product
$\sigma_{co}^{\psi} dN^h/dy$. The value of $b$ in each bin is determined
from the NA38 code [15]. Unfortunately, in $Pb$ $Pb$ collisions, the
calorimeter does
not have the same acceptance as in $SU$ and, moreover, is not located at
mid-rapidities.
Due to these differences the $E_T$ measured in $Pb$ $Pb$ has to be multiplied
by a factor
$2.35 \pm 0.15$ [16] in order to be comparable to the $E_T$ measured in $SU$.
Finally for
a $pA$ collision we take $${dN^{pA \to h} \over dy}(b) = {3 \over 2} (\bar{\nu}
+ 1)
{dN^{NN \to h^-} \over dy} \ \ \ , \eqno(11)$$ \noindent where $\bar{\nu}$ is
the
average number of collisions. From (1) and (7) we obtain the combined result of
nuclear
absorption and destruction of the $J/\psi$ via interactions with co-moving
hadrons, as
$${dN^{\psi} \over dy} (b) = \left . {dN^{\psi} \over dy} \right |_{\tau_0} (b)
\
S_1(b) \ S_2(b) \eqno(12)$$ \noindent Note that $dN^{\psi}/dy$ at $\tau_0$ is
close but
not identical to $AB$ times the corresponding value in $pp$ collisions. This is
due to
the fact that, contrary to $S_1$, $S_2 \not= 1$ for $pp$. Therefore it has to
be
determined from $$\left . {dN \over dy}\right |_{\tau_0}^{AB \to \psi} (b) = AB
{dN^{pp
\to \psi} \over dy} S_2^{pp} (b) \ \ \ . \eqno(13)$$
\noindent We can now compute the absolute yield of $J/\psi$ in any reaction -
or the
ratio $J/\psi$ over DY since, in the latter case, $S_1 = S_2 = 1$. The results,
which
depend on two parameters $\sigma_{abs}$ and $\sigma_{co}^{\psi}$, are presented
in Fig.
1 and compared with the NA38 and NA50 data. The agreement with experiment is
reasonably
good. In particular the strong suppression between $SU$ and $Pb$ $Pb$ is
obtained with no
discontinuity in the parameters. However, our $L$-dependence is somewhat too
weak
in $pA$ collisions and too strong in $SU$. It is important that the values of
the
parameters $$\sigma_{abs} = 4.1 \ {\rm mb} \qquad , \qquad \sigma_{co}^{\psi} =
0.46 \
{\rm mb} \eqno(14)$$ \noindent are very reasonable. Had we needed a much larger
value of
$\sigma_{co}^{\psi}$ our interpretation of the $J/\psi$ suppression should be
dismissed
on theoretical grounds [6]. \par

So far we have considered, besides nuclear absorption, all the destruction
channels $h +
\psi \to D + \bar{D} + X$, ... , with cross-section
$\sigma_{co}^{\psi}$. Likewise, in order to study $\psi '$ suppression we have
to
consider the channels $h + \psi ' \to D + \bar{D} + X$, ... , which do not
involve
$\psi '$ in the final state. The corresponding cross-section will be denoted
$\sigma_{co}^{\psi '}$. Due to the different geometrical sizes,
$\sigma_{co}^{\psi '}$
is larger than $\sigma_{co}^{\psi}$ at very high energies. Their difference is
even
bigger at low energies due to the dramatic differences in the energy behaviour
of these
two cross-sections near threshold [6]. With this sole extra parameter at our
disposal,
it is not possible to reproduce the $\psi '/\psi$ ratio in both $SU$ and $Pb$
$Pb$
systems. If we choose $\sigma_{co}^{\psi '}$ such as to reproduce the $SU$
data, the
result for central $Pb$ $Pb$ is an order of magnitude too low. However, in this
case
the above destruction channels are not the only relevant ones. One has also to
consider the exchange channels $$\psi + \pi \longrightarrow \psi '  + X \quad ,
\quad  \psi ' + \pi \longrightarrow \psi  + X \eqno(15)$$
\noindent with cross-sections $\sigma_{ex}^{\psi}$ and $\sigma_{ex}^{\psi '}$
respectively. Asymptotically, $\sigma_{ex}^{\psi} = \sigma_{ex}^{\psi '}$.
However, at
low energies $\sigma_{ex}^{\psi '}$ is expected to be much larger than
$\sigma_{ex}^{\psi}$ due to the different thresholds. The presence of these
channels
has little effect on the $J/\psi$ over DY ratio but it changes considerably the
$\psi
'/\psi$ one and allows to cure the problem mentioned above. Indeed, for central
$Pb$
$Pb$ collisions, when the $\psi '/\psi$ ratio becomes very small, channels (15)
produce
a feeding of $\psi '$ at the expense of $\psi$, thereby increasing the ratio
$\psi
'/\psi$. \par

Let us now discuss the combined effect of all destruction and exchange
channels.
The destruction channel for the $\psi '$ is treated in the same way as for the
$\psi$
- with $\sigma_{co}^{\psi}$ replaced by $\sigma_{co}^{\psi '}$
($\sigma_{co}^{\psi
'} > \sigma_{co}^{\psi}$). For the exchange channels (15), there is a gain of
$\psi '$
due to $\psi \to \psi '$ conversion and a loss of $\psi$ due to the inverse
reaction.
The net gain of $\psi '$ is
$$\Delta (b) = \left . {dN^{\psi} \over dy} \right |_{\tau_0} (b) \left [
\sigma_{ex}^{\psi} - \sigma_{ex}^{\psi '} R(b) \right ] G(b) \ \ell n \left |
{\tau_0 +
\Delta \tau \over \tau_0} \right | {dN^h \over dy}(b) \eqno(16)$$
\noindent where $R(b)$ is the ratio of $\psi '$ over $\psi$ rapidity densities
at time
$\tau_0$.  The net gain of $\psi$ is obviously given by the same eq. (16)
with opposite sign.  \par

Combining (1), (6) and (16) we have
$$\left . {dN^{\psi} \over dy} \right |_{\tau_0 + \Delta \tau} (b) = \left [
\left .
{dN^{\psi} \over dy} \right |_{\tau_0}(b) \exp \left [ - \sigma_{co}^{\psi}
G(b) \ell n
\left | {\tau_0 + \Delta \tau \over \tau_0} \right | {dN^h \over dy}(b) \right
] -
\Delta (b) \right ] S_1(b) \eqno(17)$$
\noindent and
$$ \left . {dN^{\psi '} \over dy} \right |_{\tau_0 + \Delta \tau} (b) = \left [
\left .
{dN^{\psi '} \over dy} \right |_{\tau_0}(b) \exp \left [ - \sigma_{co}^{\psi '}
G(b) \ell n \left | {\tau_0 + \Delta \tau \over \tau_0} \right | {dN^h \over
dy}(b)
\right ] + \Delta (b) \right ] S_1(b)  \ \ \ . \eqno(18)$$
\noindent Contrarily to (6), eqs. (17) (18), have to be solved numerically,
because, not
only ${dN^{\psi} \over dy}$ changes with increasing $\tau$, but also $R(b)$.
Therefore,
it is not possible to get a close formula at freeze-out time $\tau$ - but only
the
variation during an infinitesimal interval $\Delta \tau$. One has to solve the
problem
nu\-me\-ri\-cally, dividing the total $\ell n \tau$ interval into a very large
number of
subintervals, and using as initial condition in each subinterval the result
obtained at
the end of the previous one. \par

The results for the ratios $J/\psi$ over DY and $\psi '/\psi$ are given in
Figs. 1 and
2. We have used the following values of the parameters
$$\sigma_{abs} = 4.1 \ {\rm mb} \ , \ \sigma_{co}^{\psi} = 0.40 \ {\rm mb}
\ , \ \sigma_{co}^{\psi '} = 2.6 \ {\rm mb} \ , \ \sigma_{ex}^{\psi} =
0.1 \ {\rm mb} \ , \  \sigma^{\psi '}_{ex} = 0.65 \ {\rm mb} \eqno(19)$$
\noindent The result presented in Fig. 1 for the $J/\psi$ over $DY$ ratio is
not
changed by the introduction of the exchange channels (15) (within 1 $\%$). More
precisely, a small change in $\sigma_{co}^{\psi}$ from 0.46 (14) to 0.40 mb
(19) has
compensated for their effect. The value of $\sigma_{co}^{\psi '}$ is basically
determined from the data on $\psi '/\psi$ for SU. Finally, the value of
$\sigma_{ex}^{\psi}$ is determined in such a way to get enough feeding of $\psi
'$ from
$\psi$ in $Pb$ $Pb$. Due to the smallness of $R(b)$, our results are rather
unsensitive
to the ratio $\sigma_{ex}^{\psi '}/\sigma_{ex}^{\psi}$ and, in order to
decrease the
number of parameters, we have taken it equal to $\sigma_{co}^{\psi
'}/\sigma_{co}^{\psi} = 6.5$. (A ratio $\sigma_{ex}^{\psi '}/\sigma_{ex}^{\psi}
= 1$
with $\sigma_{ex}^{\psi} = 0.06$ also gives acceptable results). Although we
have not
attempted a best fit of the data we describe the $\psi '/\psi$ ratio reasonably
well.
In particular, we have a mild decrease of this ratio both in $pA$ and $Pb$ $Pb$
collisions and a faster decrease in $SU$. This striking feature is also present
in
the experimental data. However, our $\psi '/\psi$ ratio in $pA$ collisions
decreases
somewhat faster than the experimental one. \par

Before concluding it should be noted that the values of $E_T$ measured in $Pb$
$Pb$
collisions, relative to those measured in $SU$, are 20 to 30 $\%$ larger than
expected
from scaling in the number of participant nucleons and from Monte Carlo codes.
At present
this point is not well understood either theoretically or experimentally. If
the $E_T$
values in $Pb$ $Pb$ were to be decreased by such an amount, the values of the
ratio
$\psi '/\psi$ in $Pb$ $Pb$ would increase without spoiling the agreement with
experiment. However, the ratio $J/\psi$ over DY for $Pb$ $Pb$ collisions would
increase
(by as much as 20 $\%$ in the most central bin of $Pb$ $Pb$) as shown in Fig.
1. In this
case, the mechanism described above would not reproduce entirely the NA50 data.
\par

Note also that an important part of the effect of the co-movers comes from the
region
of $\tau$ near $\tau_0$ where the densities are very high and one can wonder
whether
such a dense system can be regarded as a hadronic one. In any case our
mechanism of
$J/\psi$ suppression is different from Debye screening. \par

In conclusion, combining nuclear absorption and final state interaction with
co-moving
hadrons, we have obtained a reasonable description of the $J/\psi$ and $\psi '$
data.
This des\-crip\-tion is better for $SU$ and $Pb$ $Pb$ collisions than for $pA$.
It has
been achieved with theoretically meaningful values of the cross-sections
involved and
without introducing any discontinuity in the $J/\psi$ or $\psi '$ survival
probabilities. \par \vskip 3 truemm

\noindent {\bf Acknowledgments} \par
It is a pleasure to thank J. P. Blaizot, M. Braun, B. Chaurand, S. Gavin, M.
Gonin, R.
Hwa, D. Kharzeev, L. Kluberg, A. Krzywicki, A. Mueller, J. Y. Ollitrault, C.
Pajares,
H. Satz, Y . Shabelski, D. Schiff and J. Tran Thanh Van for discussions.

\vfill\supereject \centerline{\bf References} \vskip 3 truemm
\baselineskip=16 pt
\item{[1]} T. Matsui and
H. Satz, Phys. Lett. {\bf B178} (1986) 416. \item{[2]} NA38 collaboration : C.
Baglin et
al, Phys. Lett. {\bf B201} (1989) 471~; Phys. Lett. {\bf B255} (1991) 459.
\item{[3]} A.
Capella, J. A. Casado, C. Pajares, A.V. Ramallo and J. Tran Thanh Van, Phys.
Lett. {\bf
B206} (1988) 354. \item{} A. Capella, C. Merino, C. Pajares, A. V. Ramallo and
J. Tran
Thanh Van, Phys. Lett. {\bf B230} (1989) 149.
\item{[4]} C. Gerschel and J. Hfner, Phys. Lett. {\bf B207} (1988) 253~; Z.
fr Phys.
{\bf C56} (1992) 71.
\item{[5]} J.P. Blaizot and J.Y. Ollitrault, Phys. Rev. {\bf D39} (1989) 232.
\item{} S. Gavin, M. Gyulassy and A. Jackson, Phys. Lett. {\bf 207} (1988) 194.
\item{} J. Ftacnik, P. Lichard and J. Pitsut, Phys. Lett. {\bf B207} (1988)
194.
\item{} R. Vogt, M. Parakash, P. Koch and T. H. Hansson, Phys. Lett. {\bf B207}
(1988)
263.
\item{} S. Gavin and R. Vogt, Nucl. Phys. {\bf B345} (1990) 1104.
\item{} S. Gavin, H. Satz, R. Thews and R. Vogt, Z. Phys. {\bf C61} (1994) 351.
\item{[6]} G. Bhanot and M. E. Peskin, Nucl. Phys. {\bf B156} (1979) 365.
\item{} A. Kaidalov and P. Volkovitsky, Phys. Rev. Lett. {\bf 69} (1992) 3155.
\item{} A. Kaidalov, Proceedings XXVIII Rencontres de Moriond (1993), ed. J.
Tran
Thanh Van. \item{}JM. Luke et al, Phys. Lett. {\bf B288} (1992) 355.
\item{} D. Kharzeev and H. Satz, Phys. Lett. {\bf B306} (1994) 155.
\item{[7]} D. Kharzeev and H. Satz, Phys. Lett. {\bf B366} (1996) 316.
\item{} B. Kopeliovich and J. Hfner, Phys. Rev. Lett. {\bf 76} (1996) 192.
\item{[8]} NA50 collaboration : P. Bordal\'o et al, Proceedings XXI Rencontres
de
Moriond (1996), ibid~; M. Gonin et al, Proceedings Quark Matter 96, to be
published in Nucl. Phys.
A. \item{[9]} J.P.
Blaizot and J.Y. Ollitrault, Proceedings Quark Matter 96, ibid and Saclay
preprint 1996.
\item{} C.Y. Wong, Proceedings Quark Matter 96, ibid.
\item{} D. Kharzeev and H. Satz, Proceedings QM'96, ibid.
\vfill \supereject
\item{[10]} Preliminary results of this work were presented by A. Capella
during the
discussion session in the QM'96 conference. For a contribution based on a
similar
approach, see S. Gavin and R. Vogt, Proceedings of QM'96 ibid.
\item{[11]} A. Capella, Phys.
Lett. {\bf B364} (1995) 175. \item{} A. Capella, A. Kaidalov, A. Kouider Akil,
C. Merino
and J. Tran Thanh Van, Z. Phys. {\bf C70} (1996) 507.
\item{[12]} P. Koch, U. Heinz, J. Pitsut, Phys. Lett. {\bf B243} (1990) 149.
\item{[13]} A. Capella and A. Krzywicki, Z. Phys. {\bf C41} (1989) 659.
\item{[14]} NA35 collaboration : G. Roland, Nucl. Phys. {\bf A566} (1994) 527c.
\item{} D. Ferenc, Proceedings XXIX Rencontres de Moriond (1994) ibid.
\item{[15]} NA38 collaboration~: C. Baglin et al, Phys. Lett. {\bf B251} (1990)
472.
\item{[16]} NA50 collaboration : private communication.
\vskip 2 truecm
\centerline{\bf Figure Captions} \vskip 5 truemm
{\parindent=1 cm
\item{\bf Fig. 1} The ratio $B_{\mu \mu} \sigma (J/\psi )/\sigma (DY)$ versus
the
interaction length $L$ in the final state for $pp$, $pA$, $SU$ and $Pb$ $Pb$
collisions. The data are from ref. [8]. The theoretical values are obtained
from eq.
(17) with the values of the parameters in (19). The same result (within 1 $\%$)
is
obtained from eq. (12) with the values of the parameters in (14). The straight
line
corresponds to nuclear absorption alone (eq. (1)), with $\sigma_{abs} = 6.2$
mb.

\vskip 3 mm

\item{\bf Fig. 2} The ratio $B_{\mu \mu} \sigma (\psi ')/B_{\mu \mu} \sigma
(J/\psi )$
versus $L$ in $pp$, $pA$, $SU$ and $Pb$ $Pb$ collisions. The data are from ref.
[8].
The theoretical values are obtained from eq. (18), with the values of the
parameters
in (19).  \vskip 3 mm

\par}

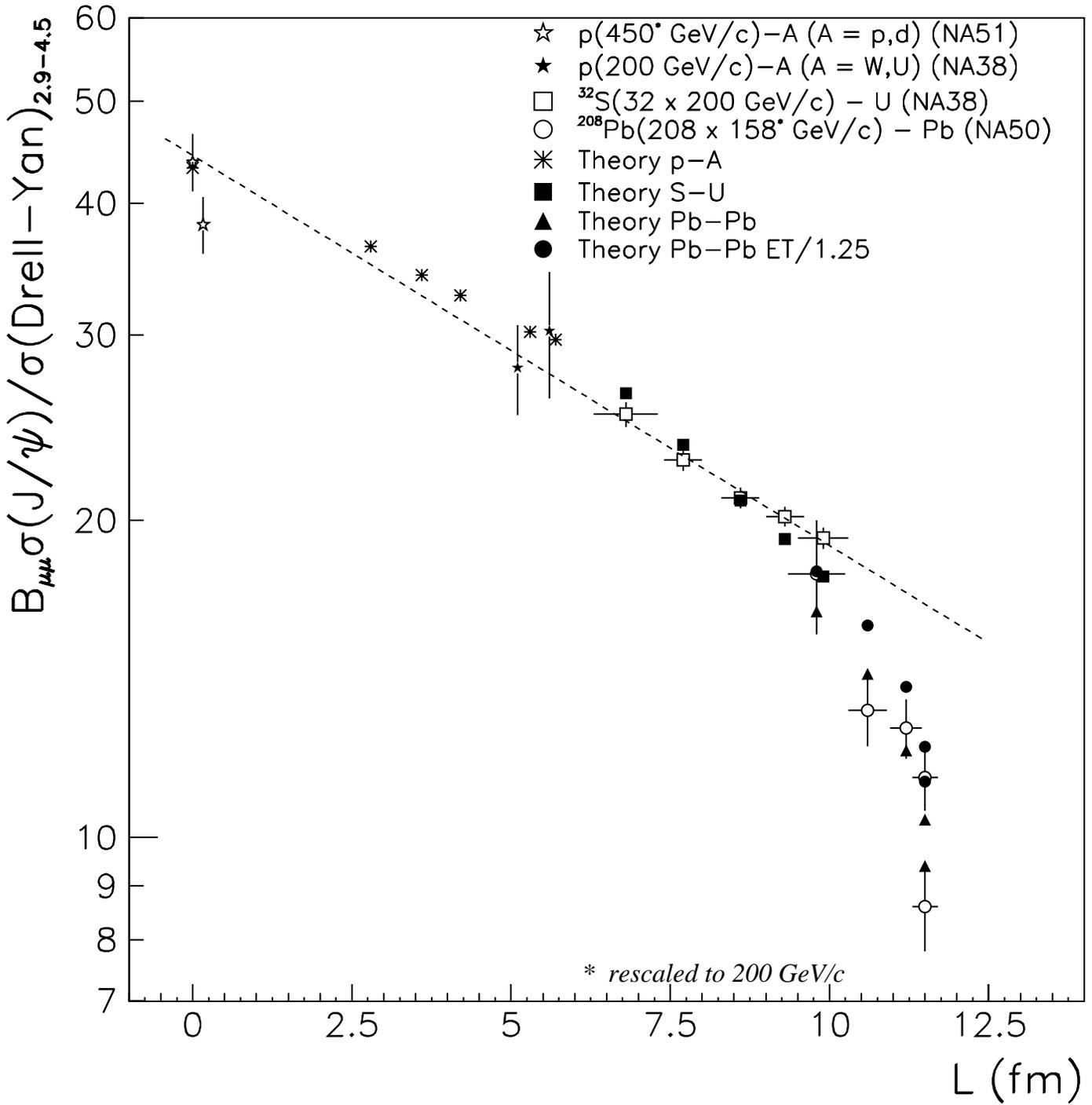
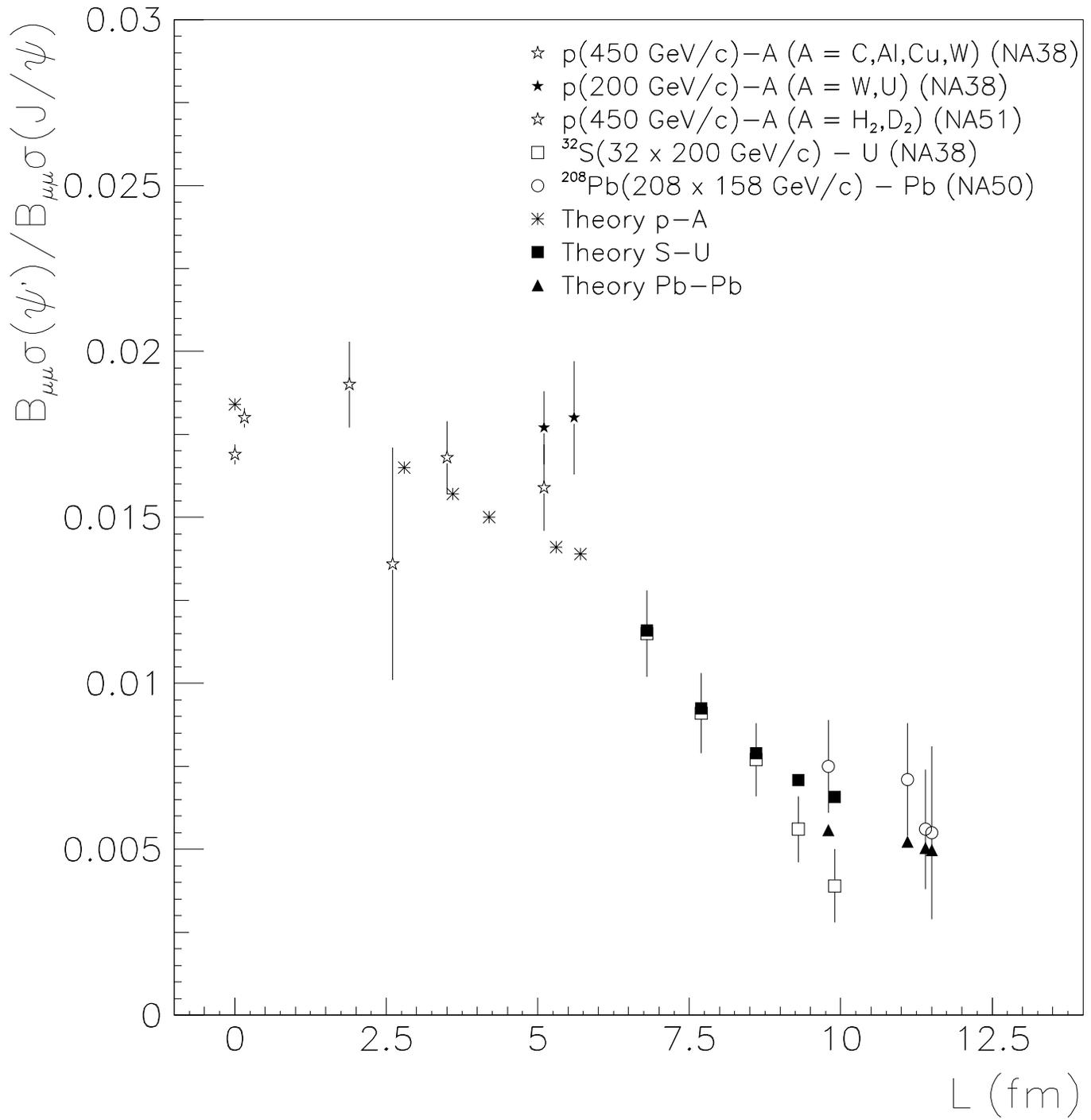
\bye